# On Verification of the Non-Generational Conjectural-Derivation of First Class constraints: HP Monopole's Field Case

K. Rasem Qandalji

*Amer Institute*
*P.O. Box 1386, Sweileh, 11910*
*JORDAN*
*E-mail*: qandalji@hotmail.com

**ABSTRACT**

In [7] we proposed a non-generational conjectural-derivation of all first class constraints (involving, only, variables compatible with canonical Poisson brackets) for "realistic" gauge (singular) field theories; and we verified the conjecture in cases of electromagnetic field, Yang-Mills fields interacting with scalar and spinor fields, and the gravitational field. Here, we will further verify our conjecture for the case of 't Hooft-Polyakov (HP) monopole's field (i.e. in the Higgs vacuum); and show that we will reproduce the results of Ref.[6], which were reached at using Dirac's standard multi-generational algorithm.

## 1. Introduction and Preliminaries

**1.1. Dirac's Standard Algorithm.**[1,2,3,4,8] In singular field theories, the conjugate momenta are not all independent and therefore, they satisfy one or more "primary" constraints of the form

$$\Phi(q,p) \approx 0. \qquad (1)$$

Eqs.(1) result from the Lagrangian being singular; where velocities cannot be solved in terms of the coordinates and momenta alone. Eqs.(1) have to be satisfied at all times since momenta's defining functions have to be satisfied at all times too; so:

$$\dot{\Phi}(q,p) \approx 0. \qquad (2)$$

Unless (2) are identically satisfied using (1), or that (2) lead to inconsistency due to Lagrangian being inconsistent; Eqs.(2) will impose new conditions on the velocities which are not yet explicitly expressible in terms of coordinates, momenta, and other un-expressible velocities alone, otherwise (2) may lead to new equations of the form (1) which again may lead to new conditions of the form (2), and so on. Eventually, we reach a stage or "generation" at which no new equations, of the form (1), can be produced. [Constraints, we get at all subsequent stages beyond the first one, are called "secondary constraints"].

If, by the final stage, we are able to express all the velocities in terms of the coordinates and the momenta alone, then, this will mean that all the constraints which we obtained so far are "second-class constraints"; and the theory will have no "gauge" degeneracy. If, on the other hand, we are not able, by the final stage, to express all the velocities as functions of the coordinates and that momenta alone, then, this indicates we have one or more "first-class constraints" that have vanishing Poisson brackets with all other constraints; which requires the introduction of gauge-fixing conditions that will remove the "gauge" degeneracy caused by these first-class constraints.

**1.2. Conjecture for Non-Generational Derivation of All First-class Constraints at Once.**
In [7] we proposed a conjectural-rule for deriving all the first-class constraints (responsible for the gauge degeneracy), of all stages, in a single step based on the generators of the gauge degeneracy in the Lagrange's formulation, provided that these first-class constraints only involve "canonical" variables [i.e., variables that do not violate the equal-time canonical Poisson bracket relations: $\{q_a(\mathbf{x},t), p_b(\mathbf{y},t)\}_{PB} = \delta_{ab}\delta(\mathbf{x}-\mathbf{y})$].

The Euler-Lagrange equations, of a gauge field theory, must satisfy certain gauge identities. These identities can be written in the form [4],[5]:

$$\int \frac{\delta S}{\delta q^a(x)} \Re_\alpha^a(x,x')d^4x \equiv 0, \qquad [\alpha]=r, \qquad (3)$$

where $r$ is the degree of gauge degeneracy, i.e. the total number of independent identities of the form (3). The quantities $\Re_\alpha^a(x,x')$, are called "gauge generators"; they are independent and can be written as local functions of space-time in the form [5]:

$$\Re_\alpha^a(x,x') = \sum_{k_0=0}\ldots\sum_{k_3=0} \Omega_{\alpha,k_0\ldots k_3}^a(q(x),\partial^\mu q(x),\ldots) \times \partial_0^{k_0}\ldots\partial_3^{k_3}\delta(x-x'), \qquad (4)$$

In Dirac's algorithm, the first-class constraints, of all stages, are associated with the theory's gauge degeneracy and therefore, equal in number to the degree of gauge freedom.

The conjecture in Ref.[7] states that: *All first-class constraints, in a given gauge theory, can be written in a form in which the conjugate momenta, $\pi(x)$'s, inherit the same symmetry identities of the Lagrange's equations [given by (3)]*. This conjecture is restricted to those "canonical" variables that do not violate the canonical Poisson bracket relations between conjugate variables. So, this conjecture claims that all first-class constraints, (for realistic gauge field theories), that involve "canonical" variables are given by the equations

$$\int \pi_a(x)\Re_\alpha^a(x,x')d^4x \approx 0, \qquad (5)$$

where, the gauge generators, $\Re_\alpha^a(x,x')$, here are the same ones we have in (3).

In Sec.2. we review the Dirac algorithm applied to the 't Hooft-Polyakov (HP) monopole's field. In Sec.3. we reproduce the first-class constraints of Sec.2. using the conjecture stated above.

## 2. Summary of Dirac's Algorithm applied to the HP Monopole's Field. [6]

The 't Hooft-Polyakov (HP) monopole model [10,11,12,13], consists of an $SO(3)$ gauge field interacting with an isovector Higgs field $\boldsymbol{\phi}$. The model's Lagrangian is:

$$\mathcal{L} = -\frac{1}{4} G_a^{\mu\nu} G_{a\mu\nu} + \frac{1}{2} \mathcal{D}^\mu \boldsymbol{\phi} \cdot \mathcal{D}_\mu \boldsymbol{\phi} - V(\boldsymbol{\phi}),$$

where $\boldsymbol{\phi} = (\phi_1, \phi_2, \phi_3)$, and $V(\boldsymbol{\phi}) = \frac{1}{4}\lambda(\phi_1^2 + \phi_2^2 + \phi_3^2 - a^2)^2$. $G_a^{\mu\nu}$, is the gauge field strength: $G_a^{\mu\nu} = \partial^\mu W_a^\nu - \partial^\nu W_a^\mu - e\varepsilon_{abc} W_b^\mu W_c^\nu$, where $W_a^\mu$ is the gauge potential.

The model's Lagrangian full symmetry Group $SO(3)$, generated by $T_a$'s, is spontaneously broken, by the Higgs Vacuum (defined below), down to $SO(2)$ ($\simeq U(1)$), generated by $\frac{\boldsymbol{\phi} \cdot \mathbf{T}}{a}$.

The model's non-singular extended solution looks, at large distances, like a Dirac monopole. The monopole's energy finiteness implies that there is some radius $r_0$ such that for $r \geq r_0$ we have, to a good approximation:

$$\mathcal{D}^\mu \boldsymbol{\phi} \equiv \partial^\mu \boldsymbol{\phi} - e \mathbf{W}^\mu \times \boldsymbol{\phi} = 0, \tag{6}$$

and,
$$\phi_1^2 + \phi_2^2 + \phi_3^2 - a^2 = 0, \quad (\Rightarrow V(\boldsymbol{\phi}) = 0). \tag{7}$$

Regions of space-time, where the above two equations are satisfied, constitute the "Higgs Vacuum". The general form of $\mathbf{W}^\mu$ in the Higgs Vacuum is [14]:

$$\mathbf{W}^\mu = \frac{1}{a^2 e} \boldsymbol{\phi} \times \partial^\mu \boldsymbol{\phi} + \frac{1}{a} \boldsymbol{\phi} A^\mu, \tag{8}$$

where $A^\mu$ is arbitrary. It follows that:

$$\mathbf{G}^{\mu\nu} = \frac{1}{a} \boldsymbol{\phi} F^{\mu\nu} \; ; \; [\text{where, } F^{\mu\nu} = \frac{1}{a^3 e} \boldsymbol{\phi} \cdot (\partial^\mu \boldsymbol{\phi} \times \partial^\nu \boldsymbol{\phi}) + \partial^\mu A^\nu - \partial^\nu A^\mu ]. \tag{9}$$

So in "Higgs vacuum": $\mathcal{L} = -\frac{1}{4} G_a^{\mu\nu} G_{a\mu\nu}$; and on account of (7,9) we get,

$$\mathcal{L} = -\frac{1}{4} F^{\mu\nu} F_{\mu\nu}. \tag{10}$$

In Higgs vacuum region, we have the conjugate momenta of the dynamical coordinates $A^\eta(\mathbf{x})$'s and $\phi_i(\mathbf{x})$'s, given by [6,8]:

$$\Pi_\eta(x) \equiv \frac{\partial \mathcal{L}}{\partial \dot{A}^\eta(x)} = \frac{\varepsilon_{rst}}{a^3 e}\phi_r \partial_\eta \phi_s \partial_0 \phi_t + \partial_\eta A_0 - \partial_0 A_\eta = \begin{cases} 0, & \text{for } \eta = 0 \\ F_{i0}, & \text{for } \eta = i = 1,2,3 \end{cases}, \quad \text{(11a)}$$

$$\pi_m(x) \equiv \frac{\partial \mathcal{L}}{\partial \dot{\phi}_m(x)} = \frac{\varepsilon_{ijm}}{a^3 e}\phi_i \partial^k \phi_j \left(\frac{\varepsilon_{rst}}{a^3 e}\phi_r \partial_0 \phi_s \partial_k \phi_t + \partial_0 A_k - \partial_k A_0\right). \quad \text{(11b)}$$

As for the Dirac quantization of the HP monopole's field (i.e. in the Higgs vacuum region), the details are given in refs.[6]. Now, we list all the constraints (of all generations and classes), we arrived at using the standard Dirac algorithm; the complete set of constraints (in the axial gauge), and including the gauge-fixing ones, are [6]:

$$\zeta_1 \equiv \phi_2 \Phi_1 - \phi_1 \Phi_2 - \frac{\alpha_3}{2}\chi \approx 0,$$

$$\zeta_2 \equiv \phi_3 \Phi_2 - \phi_2 \Phi_3 - \frac{\alpha_1}{2}\chi \approx 0,$$

$$\zeta_3 \equiv \frac{1}{2a^2}(\phi_1 \Phi_1 + \phi_2 \Phi_2 + \phi_3 \Phi_3) \approx 0,$$

$$\zeta_4 \equiv \chi = \phi_1^2 + \phi_2^2 + \phi_3^2 - a^2 \approx 0, \text{[which is also a strong equation, see (7);}$$
$$\text{being strong also gives: } \sum_i \phi_i \partial^\mu \phi_i = 0 \text{ ]}$$

$$\zeta_5 \equiv \partial^i \Pi_i \approx 0,$$

$$\zeta_6 \equiv \frac{1}{ae}(\phi_2 \partial^3 \phi_1 - \phi_1 \partial^3 \phi_2) - A^3 \phi_3 \approx 0,$$

$$\zeta_7 \equiv \frac{1}{ae}(\phi_3 \partial^3 \phi_2 - \phi_2 \partial^3 \phi_3) - A^3 \phi_1 \approx 0,$$

$$\zeta_8 \equiv A^3 \approx 0,$$

(12a)

where,

$$\Phi_m \equiv \pi_m + \frac{\varepsilon_{ijm}}{a^3 e}\phi_i \partial^k \phi_j \Pi_k; \text{ and } \alpha_k \equiv \frac{3}{a^3 e}\Pi_j \partial^j \phi_k. \quad \text{(12b)}$$

$\zeta_1, \zeta_2$ are primary first-class constraints; $\zeta_5$ is secondary first-class; $\zeta_3, \zeta_4$ are primary second-class (with, following [6], $\zeta_4$ can be treated also as "strong" equation when needed); and $\zeta_6, \zeta_7, \zeta_8$ are gauge-fixing conditions that lift the degeneracy of the Hamiltonian associated with the first-class constraints and equal in number to them.

**Note.** Using Eq.(11a), we find yet another first-class constraint, namely,

$$\zeta_0 \equiv \Pi_0 \approx 0. \quad \text{(12c)}$$

It violates the canonical Poisson bracket relation and hence, leads to a contradiction upon passing to quantum theory. We, therefore, follow Dirac [2] in restricting $\Pi_0(x)$ and (gauge-fixing) $A^0(x)$ to "zero" at all times; i.e. we discard the $A^0$ degree of freedom, which cease to have any physical interest anymore.

## 3. Derivation of First-Class Constraints Using the Conjectured-Rule.

In this section we will use the rule in (5) to reproduce the first-class constraints $\zeta_1, \zeta_2, \zeta_5$. On the other hand, our rule (5) doesn't apply to the constraint, $\zeta_0 \equiv \Pi_0 \approx 0$, since the vanishing of $\Pi_0$ violates its canonical Poisson bracket (CPB) with $A^0$. (In Sec.1.2, we already mentioned that compatibility with the CPB was required for the application of rule (5).) Using Eqs.(9,10), we get the Euler-Lagrange equations:

$$\frac{\delta S}{\delta \phi_m(x)} \equiv \frac{\partial \mathcal{L}}{\partial \phi_m(x)} - \partial^\sigma \frac{\partial \mathcal{L}}{\partial \partial^\sigma \phi_m(x)} = -\frac{\varepsilon_{mjk}}{a^3 e}\left(\frac{3}{2} F_{\mu\nu} \partial^\mu \phi_j \partial^\nu \phi_k - \phi_j \partial^\mu \phi_k \partial^\nu F_{\mu\nu}\right), \qquad (13a)$$

[where $F_{\mu\nu}$ is given in (9) as: $F^{\mu\nu} = \frac{1}{a^3 e}\boldsymbol{\phi} \cdot (\partial^\mu \boldsymbol{\phi} \times \partial^\nu \boldsymbol{\phi}) + \partial^\mu A^\nu - \partial^\nu A^\mu$],

and, $$\frac{\delta S}{\delta A^\nu(x)} \equiv \frac{\partial \mathcal{L}}{\partial A^\nu(x)} - \partial^\mu \frac{\partial \mathcal{L}}{\partial \partial^\mu A^\nu(x)} = \partial^\mu F_{\mu\nu}. \qquad (13b)$$

Now, on account of Eqs.(13a, 13b) and the "strong" equation (7), that defines the Higgs vacuum region, (namely, $\phi_j \phi_j - a = 0 \Leftrightarrow \phi_j \partial^\mu \phi_j = 0$); we have,

$$\varepsilon_{kmn} \phi_m \frac{\delta S}{\delta \phi_n} = -\frac{1}{ae}\partial^\mu \phi_k \partial^\nu F_{\mu\nu} = \frac{1}{ae}\partial^\mu \phi_k \frac{\delta S}{\delta A^\mu},$$

that gives the three identities:

$$\varepsilon_{kmn}\phi_m \frac{\delta S}{\delta \phi_n} - \frac{1}{ae}\partial^\mu \phi_k \frac{\delta S}{\delta A^\mu} \equiv 0, \qquad k = 1, 2, 3. \qquad (14)$$

These identities are not all independent; using Eq.(7) ($\phi_j \phi_j - a = 0 \Leftrightarrow \phi_j \partial^\mu \phi_j = 0$), we get

$$\phi_k \left(\varepsilon_{kmn}\phi_m \frac{\delta S}{\delta \phi_n} - \frac{1}{ae}\partial^\mu \phi_k \frac{\delta S}{\delta A^\mu}\right) \equiv 0.$$

So we pick from (14) two "independent" identities, say: $\Im_1, \Im_2$, corresponding to, $k = 3, 1$, respectively:

$$\mathfrak{I}_1 \equiv \varepsilon_{3mn}\phi_m \frac{\delta S}{\delta \phi_n} - \frac{1}{ae}\partial^\mu \phi_3 \frac{\delta S}{\delta A^\mu} \equiv 0,$$

$$\mathfrak{I}_2 \equiv \varepsilon_{1mn}\phi_m \frac{\delta S}{\delta \phi_n} - \frac{1}{ae}\partial^\mu \phi_1 \frac{\delta S}{\delta A^\mu} \equiv 0.$$

**(15a)**

Our last independent identity will be derived from (13b) due to anti-symmetry of $F_{\mu\nu}$:

$$\mathfrak{I}_3 \equiv \partial^\nu \frac{\delta S}{\delta A^\nu(x)} = \partial^\nu \partial^\mu F_{\mu\nu} \equiv 0. \tag{15b}$$

$\mathfrak{I}_1, \mathfrak{I}_2, \mathfrak{I}_3$ can be put in the form (3):

$$\mathfrak{I}_i(x) = \int \frac{\delta S}{\delta \phi_m(x')} \mathfrak{R}_i^m(x',x) d^4x' + \int \frac{\delta S}{\delta A^\mu(x')} \mathfrak{R}_i^\mu(x',x) d^4x' \equiv 0, \quad i = 1,2,3,$$

with,

$$\mathfrak{R}_1^m(x',x) = \varepsilon_{3nm}\phi_n(x')\delta(x'-x)$$

$$\mathfrak{R}_1^\mu(x',x) = -\frac{1}{ae}\partial^{\mu'}\phi_3(x')\delta(x'-x)$$

$$\mathfrak{R}_2^m(x',x) = \varepsilon_{1nm}\phi_n(x')\delta(x'-x)$$

$$\mathfrak{R}_2^\mu(x',x) = -\frac{1}{ae}\partial^{\mu'}\phi_1(x')\delta(x'-x)$$

$$\mathfrak{R}_3^m(x',x) = 0$$

$$\mathfrak{R}_3^\mu(x',x) = -\partial^{\mu'}\delta(x'-x).$$

**(16)**

Substituting (16) into (5):

$$\mathfrak{D}'_i(x) \equiv \int \pi_m(x')\mathfrak{R}_i^m(x',x) d^4x' + \int \Pi_\mu(x')\mathfrak{R}_i^\mu(x',x) d^4x' \approx 0, \quad i = 1,2,3. \tag{17a}$$

As was said above following Eq.(12c), (i.e., following [2],[6]); we discard the $A_0$ degree of freedom by setting, $A_0(x) = \Pi_0(x) = 0$, at all times; so, Eq.(17a) will reduce to

$$\mathfrak{D}_i(x) \equiv \int \pi_m(x')\mathfrak{R}_i^m(x',x) d^4x' + \int \Pi_j(x')\mathfrak{R}_i^j(x',x) d^4x' \approx 0, \quad i = 1,2,3; \tag{17b}$$

or, explicitly, we write
$$\mathfrak{D}_1 \equiv \varepsilon_{3ik}\phi_i \pi_k - \frac{1}{ae}\Pi_j \partial^j \phi_3$$
$$\mathfrak{D}_2 \equiv \varepsilon_{1ik}\phi_i \pi_k - \frac{1}{ae}\Pi_j \partial^j \phi_1 \tag{17c}$$
$$\mathfrak{D}_3 \equiv \partial^j \Pi_j .$$

Using (12a,b), and that $\zeta_4 \equiv \chi \approx 0$ is also a strong equation, i.e., $\phi_i \partial^\mu \phi_i = 0$; we rewrite $\mathfrak{D}_1$ as

$$\begin{aligned}
\mathfrak{D}_1 &\equiv \varepsilon_{3mn}\phi_m \pi_n - \frac{1}{ae}\Pi_j \partial^j \phi_3 \\
&= \varepsilon_{3mn}\phi_m \pi_n - \frac{1}{a^3 e}(\phi_i \phi_i)\partial^j \phi_3 \Pi_j \\
&= \varepsilon_{3mn}\phi_m \pi_n - \frac{1}{a^3 e}\delta_{h3}\delta_{im}\phi_i \phi_m \partial^j \phi_h \Pi_j + \frac{1}{a^3 e}\delta_{i3}\delta_{hm}\phi_i \phi_m \partial^j \phi_h \Pi_j \\
&= \varepsilon_{3mn}\phi_m \pi_n + \frac{\varepsilon_{3mn}\varepsilon_{ihn}}{a^3 e}\phi_m \phi_i \partial^j \phi_h \Pi_j \\
&= \varepsilon_{3mn}\phi_m \Phi_n \quad ,
\end{aligned}$$

(18a)

[where, the third term we added in third equality vanishes on account of, $\phi_i \partial^\mu \phi_i = 0$.]

Similarly, we obtain,

$$\mathfrak{D}_2 = \varepsilon_{1mn}\phi_m \Phi_n \tag{18b}$$

Comparing (18a,b) with (12a), we find

$$\mathfrak{D}_1 = -\zeta_1 \text{ and } \mathfrak{D}_2 = -\zeta_2, \tag{19a}$$

where both equalities hold up to an additional term in each case that vanishes "strongly" (in Higgs vacuum); namely, the additional term is proportional to $\zeta_4 \equiv \chi$.

Finally, we see from (12a) and (17c) that:

$$\mathfrak{D}_3 = \zeta_5. \tag{19b}$$

This concludes the derivation of all first-class constraints, for the HP monopole's field, using the proposed conjectural-rule (5); provided that the variables involved in these constraints are compatible with the canonical Poisson bracket relations.

## 4. Conclusion

It was shown in Sec.3. that the conjecture (5) does work in reproducing all the first-class constraints (involving only variables compatible with canonical Poisson bracket relations) for the 't Hooft-Polyakov monopole's field and matched those derived using the standard Dirac's method in Ref.[6]. The conjecture (5) was verified in Ref.[7] for electromagnetic field, Yang-Mills fields interacting with spinor and scalar fields and the gravitational field. Our work here added further to the support of this conjecture when applied to various physical fields of nature.

## 5. Acknowledgment

I thank the Ilfat & Bah. Foundation (ed'Oreen, Btouratij) for its continuous support. I thank Professor Sudarshan for reading Sec.3.